\documentclass{iaus}
\usepackage{graphicx,epsfig}

\title[A Supergiants] 
{A Supergiants}

\author[Przybilla et al.]   
{N. Przybilla$^1$,
F. Bresolin$^2$,
K. Butler$^3$,
A. Kaufer$^4$,
R.P. Kudritzki$^2$\break
\and K.A. Venn$^5$}

\affiliation{$^1$Dr. Remeis-Sternwarte Bamberg, Sternwartstrasse 7, D-96049
Bamberg, Germany\\[\affilskip] 
$^2$Institute for Astronomy, 2680 Woodlawn Drive, Honolulu, HI 96822,
USA\\[\affilskip]
$^3$Universit\"ats-Sternwarte M\"unchen, Scheinerstrasse 1, D-81679
M\"unchen, Germany\\[\affilskip]
$^4$European Southern Observatory, Alonso de Cordova 3107, Casilla 19001, 
Santiago 19, Chile\\[\affilskip]
$^5$Macalester College, 1600 Grand Avenue, Saint Paul, MN 55105, USA}

\pubyear{2004}
\volume{224}  
\pagerange{1--8}
\date{\today}
\jname{The A-Star Puzzle}
\editors{J.\,Zverko, W.W.\,Weiss, J.\,\v{Z}i\v{z}\v{n}ovsk\'{y}, \& S.J.\,Adelman, eds.}
\begin{document}

\maketitle

\begin{abstract}
A-type supergiants are primary targets for quantitative spectroscopy of
individual stars in nearby galaxies because of their intrinsic brightness.
An overview is given on the non-LTE techniques required for their
analysis. Applications
concentrate on placing observational constraints on evolutionary models for
massive stars and their host galaxies by detailed abundance analyses.
Results from high-resolution studies of A-type supergiants in Local Group
galaxies and from intermediate-resolution multi-object spectroscopy of supergiants 
far beyond the Local Group are summarised.
\end{abstract}

\firstsection 
\section{Introduction}
Massive stars of $\sim$8--40\,M$_{\odot}$ cross the A-star regime of the 
HRD during their post-main sequence evolution. 
Being supergiants at that time they are characterised by extended 
atmospheres, stellar radii measuring several tens to a few hundred
R$_{\odot}$, and immense luminosities, on the order of 10$^4$ to several
10$^5$\,L$_{\odot}$. 
The enormous intrinsic brightness, in coincidence with low
bolometric corrections, makes BA-type supergiants primary targets for the
young field of extragalactic stellar astronomy. Using 8m-class telescopes
these objects become accessible to high-resolution spectroscopy in the
galaxies of the Local Group, and to medium-resolution spectroscopy out to
distances of several~Mpc. 

This allows observational constraints to be placed on stellar evolution 
in a variety of galactic environments, 
in particular, on the effects of metallicity and rotation on stellar 
mass loss and the efficiency of chemical mixing.
Moreover, important contributions can be made for the study of the
galactochemical evolution of the host galaxies through the determination of 
present-day abundance patterns and gradients. BA-type supergiants can
help us to verify classical studies of nebulae and extend the elemental inventory 
to iron-group and s-\,\&\,r-process species. Finally, they can act as
extragalactic distance indicators, via application of the wind momentum--luminosity 
and flux-weighted gravity--luminosity relationships (WLR: Kudritzki \& Puls~2000;
FGLR: Kudritzki \etal~2003).

In order to exploit the full potential, a few complications have to be
overcome in model atmosphere analyses. The high luminosities drive stellar
winds, to be solved in a hydrodynamical approach, and low atmospheric densities 
facilitate departures from LTE, which require a simultaneous solution of 
radiative transfer and statistical equilibrium (e.g. Kub\'at \&
Kor\v{c}\'akov\'a; Krti\v{c}ka \& Kub\'at, these proceedings). In the following, we will
concentrate on the latter aspect, as this allows us to draw important conclusions 
for studies of `normal' A-stars as well, before discussing recent highlights
from the quantitative spectroscopy of extragalactic A-type supergiants.

\section{Quantitative Spectroscopy of A-type supergiants}
\begin{figure}
\includegraphics[width=.98\linewidth]{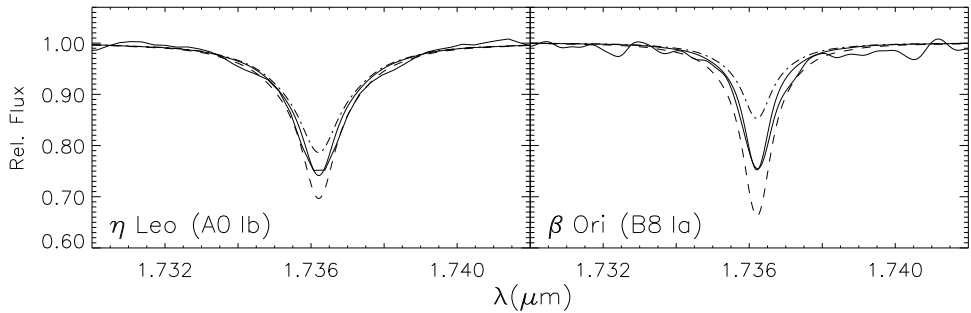}
\caption{The observed H\,{\sc i} $n$\,$=$\,4--11 transition (thick line)
in two supergiants is well reproduced by the recommended H\,{\sc i} non-LTE model
atom of Przybilla \& Butler (2004, thin line), which accounts for accurate 
electron collision data from quantum-mechanical {\em ab-initio} computations for
excitation processes. For comparison, synthetic spectra from computations
using the \cite{Johnson72} approximation (dashed) and assuming LTE (dashed-dotted 
line) are also shown. The computations are performed for
stellar parameters derived from the analysis of the visual spectra
($\eta$\,Leo: $T_{\rm eff}$\,$=$\,9600\,K, $\log g$\,=2.00; $\beta$\,Ori:
$T_{\rm eff}$\,$=$\,12000\,K, $\log g$\,=1.75). An analogous comparison in
main sequence stars like Vega indicates that non-LTE departures are
underestimated when using the \cite{MHA75} approximation for electron
collision rates.}
\label{brackett}
\end{figure}

A-type supergiants were rediscovered as tools for astrophysics in the
seminal work of \cite{Venn95}, where modern LTE model atmosphere techniques 
were shown to suffice for their quantitative analysis, using spectroscopic
indicators -- ionization equilibria and Stark-broadened hydrogen lines --
for the stellar parameter determination (cf. Y\"uce; Tanr\i{verdi}
\etal{,} these proceedings). In the
following, the focus shifted from bright, though typically less luminous
(LC Ib and Iab) Galactic objects towards
supergiants in other galaxies of the Local Group, where only the more
luminous stars are accessible to high-resolution spectroscopy using
the currently available telescopes. Finally, when meeting the challenge
of quantitative spectroscopy of supergiants beyond the Local Group only
objects near the the Eddington limit are accessible with 
present-day~instrumentation. 

The progress on the observational side initiated a
reinvestigation of the analyses techniques. Classical line-blanketed 
LTE atmospheres still turn out to be the best choice for studies of
high-luminosity objects at present, however only in combination with massive non-LTE
line-formation for the modelling of the photospheric spectrum, i.e. a hybrid 
non-LTE approach (Przybilla 2002). This is facilitated by the fact that the
main atmospheric constituents (H, He) and the important metal opacities stay
close to detailed equilibrium. In the following we discuss the major results
from these investigations.

We begin with the most basic element, hydrogen, for which the effects of non-LTE
departures in early-type stars were investigated more than three decades ago 
(e.g. Auer \& Mihalas 1969a,b). Surprisingly, present-day modelling of the
hydrogen spectrum fails in reproducing the observed Paschen,
Brackett and Pfund lines, both in LTE and non-LTE, though good agreement is
obtained for the Balmer lines. All early-type stars
are affected, with the discrepancies reaching a maximum in the A-type
supergiants, where a mismatch in the line strengths by factors up to 2--3
are found, see Fig.~\ref{brackett}. The spectral features in the 
visual and IR are consistently reproduced only when commonly-used 
approximation formulae for the evaluation of electron-collision excitation rates 
are dropped in favour of exact data from quantum-mechanical {\em ab-initio} 
computations (Przybilla \& Butler~2004). This is because of a strong
sensitivity of the line source-function to non-LTE departures in the Rayleigh-Jeans 
limit, resulting in an amplification of non-LTE effects in the IR.

\begin{figure}
\includegraphics[width=.98\linewidth]{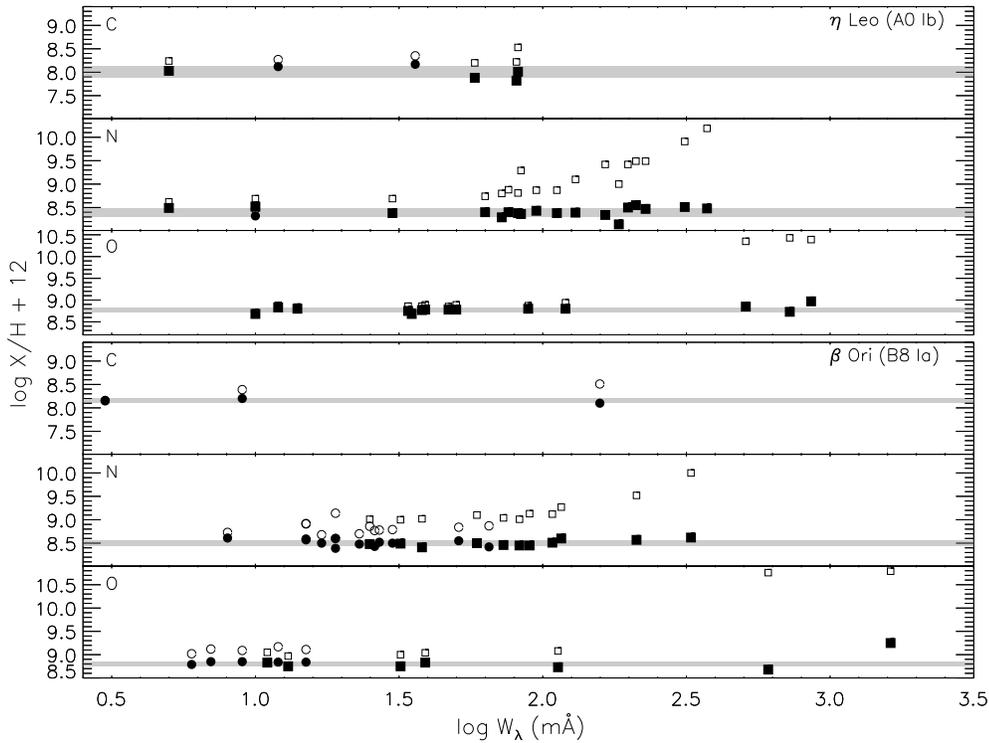}
\caption{Elemental abundances in two Galactic supergiants from individual
spectral lines of CNO plotted as a function of equivalent width: non-LTE
(solid) and LTE results (open symbols) for neutral (boxes) and
singly-ionized species. The grey bands cover the 1$\sigma$-uncertainty ranges
around the mean values: proper non-LTE calculations reduce the
line-to-line scatter and remove systematic trends.
Note that even weak lines can show considerable departures from LTE.}
\label{cnoabus}
\end{figure}

Carbon, nitrogen and oxygen are the most abundant metals. All the observed
spectral lines in the visual/near-IR originate from high-excitation, 
(quasi-)metastable levels that favour departures from LTE. Comprehensive
non-LTE model atoms, accounting for more sophisticated atomic data than
previously possible (Przybilla \etal~2000, 2001; Przybilla \& Butler~2001)
allow us to reproduce the entire observed CNO spectra to an unprecedented
degree of accuracy, see Fig.~\ref{cnoabus}. In particular, non-LTE
abundance analyses remove systematic trends that trouble the LTE approach, and help to
reduce the statistical uncertainties from the line-to-line scatter 
of typically $\sim$0.2\,dex in the literature down to better than 0.1\,dex.  
Contrary to common assumptions, significant non-LTE abundance corrections by
$\sim$0.3\,dex can occur even in the weak line limit.

\begin{figure}
\includegraphics[width=.98\linewidth]{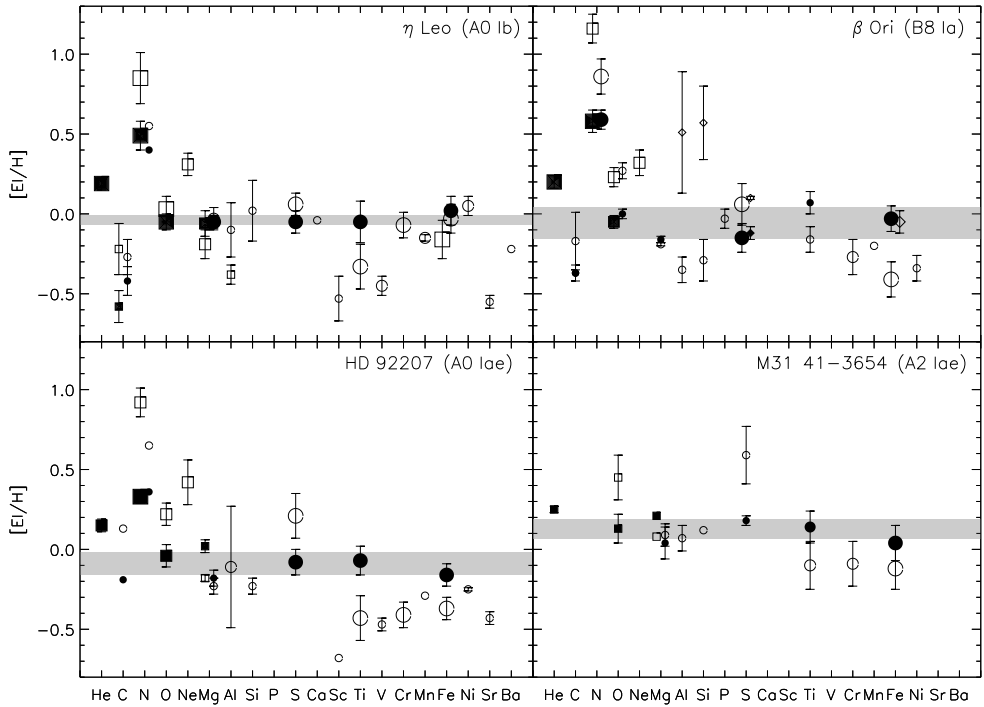}
 \caption{Abundance patterns for four Galactic and M31 supergiants, relative to
the solar standard (Grevesse \& Sauval~1998) on a logarithmic scale. Symbol
designation as in Fig.~\ref{cnoabus}, with the addition of double-ionized
species (diamonds). The symbol size codes the number of spectral lines analysed. Error
bars represent 1$\sigma$-uncertainties from the line-to-line scatter and the
grey shaded areas mark the deduced stellar metallicity within
1$\sigma$-errors. The non-LTE computations reveal a striking similarity to
the solar abundance distribution, except for the light elements which have
been affected by mixing with nuclear-processed matter.}
\label{abus}
\end{figure}

The next step is to broaden our discussion towards a comprehensive study of
the entire spectra of BA-type supergiants. Examples are shown in
Fig.~\ref{abus}, where results from non-LTE and LTE abundance analyses of
primarily weak lines are
compared to the solar standard. From this we conclude that the non-LTE analysis 
reveals a striking similarity of the Galactic supergiant abundance patterns to the solar 
abundance distribution. This is also found for the M31 object, at slightly
higher average metallicity. Fewer chemical species are accessible in this case because of 
a more restricted spectral coverage and lower S/N of the observations. 
The light elements He, C and N
show marked deviations which are interpreted as mixing of the atmospheric
layers with nuclear-processed matter, qualitatively in good agreement with
the predictions of the most recent models of massive star evolution (e.g. Maeder
\& Meynet~2000). Note in particular that non-LTE calculations can bring
several ionization equilibria simultaneously into agreement, thus putting
very tight constraints on the stellar parameters. LTE analyses on
the other hand produce a large scatter of the individual elemental
abundances, and result in increased uncertainties for the different species.
They can even suggest $\alpha$-enhancement for the more luminous
objects, i.e. apparent overabundances of the $\alpha$-elements, in
coincidence with underabundant iron-group elements. This occurs because of
selective non-LTE effects, which favour non-LTE line-strengthening in the
$\alpha$-elements in analogy to CNO.
On the other hand, iron-group elements experience
non-LTE line-weakening, because they are characterised
by a plethora of energetically-close levels easily coupled via collisions
that are collectively subject to non-LTE overionization. Ignoring these non-LTE effects
can introduce systematic errors to abundance analyses of more luminous supergiants of
typically $\sim$0.3\,dex. Though the astrophysically most
important elements are covered by our non-LTE computations, an extension to 
other chemical species is desirable. However, in many cases a lack of the
required atomic data prevents such efforts at present.

Finally, we like to emphasise that using A-type supergiants as 
testbeds for the study of non-LTE effects offers a unique opportunity to improve 
stellar analysis techniques for other classes of stars in general. Due to 
the universality of atomic properties, sets of reference atomic data should
be compiled and verified under the most extreme conditions.

\section{A-type Supergiants in the Local Group}
\begin{figure}
\vspace{.5cm}
\rule{2.7cm}{0cm}\includegraphics[width=.6\linewidth]{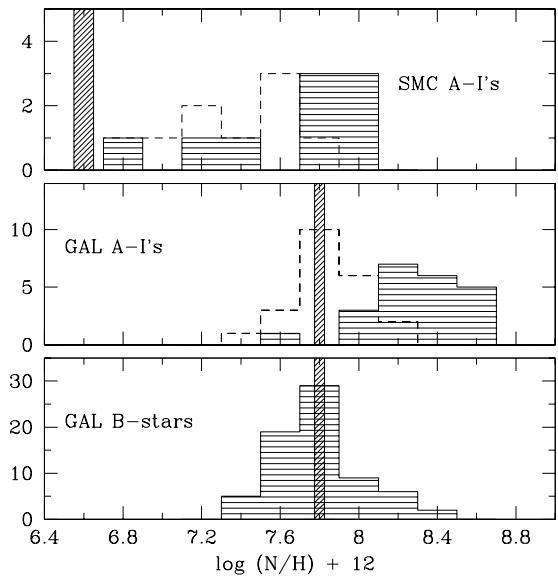}\\[1.5cm]
\caption{Histogram of nitrogen abundances in Galactic and SMC A-type
supergiants, as obtained from an improved non-LTE reanalysis (Venn \&
Przybilla 2003) of the Venn (1999, dashed line) data, and of
Galactic B-stars (Gies \& Lambert 1992; Cunha \& Lambert 1994), 
their main-sequence progenitors. The initial solar and SMC ISM
nitrogen abundances are indicated.}
\label{mixing}
\end{figure}

The different galactic environments of the individual Local Group members
offer unique opportunities to study the influence of metallicity on the
evolution of massive stars. A comparison of Galactic and SMC A-type
supergiant abundances (Venn~1999; Venn \& Przybilla~2003) suggests that most of the
objects have undergone substantial mixing with CN-cycled material, see
Fig.~\ref{mixing}. The efficiency of rotational mixing appears to correlate
with metallicity since the SMC stars (at 0.2$\times$solar metallicity) show
larger nitrogen enrichments. This is again in good qualitative agreement
with the predictions of the latest models of stellar evolution. The effect
is a consequence of the reduced metal-line opacity which gives more
compact objects and lower mass-loss rates, such that angular momentum losses
are considerably reduced, which enables the mixing mechanisms (meridional
circulation, shear instabilities) to retain their efficiency. 
A predicted correlation of mixing efficiency with stellar mass is not
verified in this study. Note however, that the sample objects in the SMC 
are on average more massive than the Galactic supergiants. 

\begin{figure}
\rule{1.7cm}{0cm}\includegraphics[width=.75\linewidth]{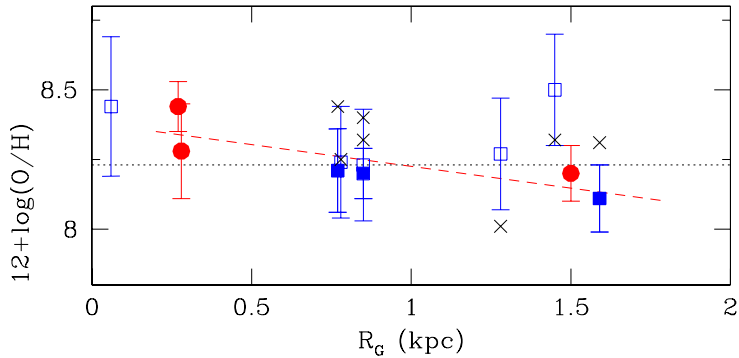}\\[-5.4cm]
\caption{Stellar (Venn \etal\ 2001, filled circles) and nebular oxygen abundances as a
function of galactocentric distance in NGC\,6822
suggest a galactic abundance gradient of $\sim$\,$-$0.1\,dex/kpc (linear
least-squares fit: dashed line). H\,{\sc ii} region abundances are
from Pagel \etal\ (1980, filled squares: abundances derived from O\,{\sc ii} and
O\,{\sc iii} lines, hollow squares: from O\,{\sc ii} alone). The mean
nebular abundance is indicated by the dotted line. Results from a
redetermination of the nebular abundances by \cite{Pilyugin01} are also
displayed (St. Andrews crosses). Note that Pilyugin's mean abundance is
$\sim$0.1\,dex higher.}
\label{gradient}
\end{figure}

The emission spectra of nebulae have been the primary sources for 
chemical abundances in extragalactic systems beyond the Magellanic Clouds
until recently. Despite a widespread use for extragalactic applications,
abundance determination techniques for H\,{\sc ii} regions (and planetary
nebulae) are still subject to a number of inherent problems (see e.g.
Stasi\'nska~2004). Abundances from individual stars in these galaxies open
up the opportunity to verify such studies with independent and well
understood indicators. Results from a comparison of nebular and stellar
oxygen abundances in the dIrr galaxy NGC\,6822 are displayed in
Fig.~\ref{gradient}. Both match within their mutual uncertainties in this
low-metallicity case.
Unexpected comes the indication of an abundance gradient,
if a sub-set consisting of the stellar and the more reliable nebular data (i.e.
observations showing both, O\,{\sc ii} and O\,{\sc iii}) is considered, 
which, if confirmed by additional measurements, could provide strong
constraints on mixing timescales for galactochemical~evolution.

\begin{figure}
\rule{1.7cm}{0cm}\includegraphics[width=.75\linewidth]{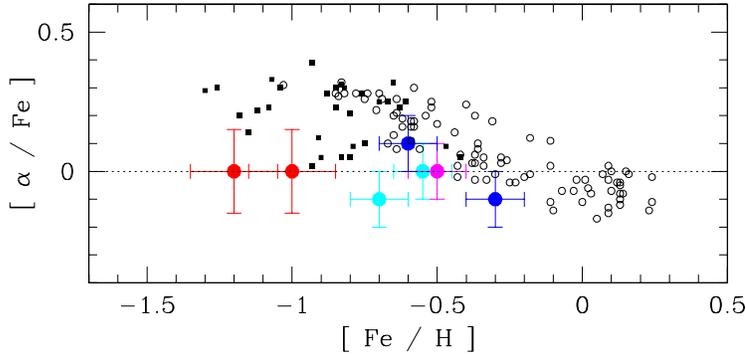}\\[-5.4cm]
\caption{Stellar $[\alpha$/Fe$]$ versus $[$Fe/H$]$ in dwarf irregular galaxies (solid
circles), from lowest to highest metallicity: Sex A, GR8, WLM, SMC (average), 
WLM (2nd object), NGC\,6822 (average of 3 objects), LMC (average). For comparison,
Galactic disk stares (open circles, Edvardsson \etal\ 1993) and metal-rich
halo stars (filled squares, Nissen \& Schuster 1997) are also shown.}
\label{dirrabundances}
\end{figure}

While nebulae can provide abundances for a variety of light
elements, iron-group and s- \& r-process elements are accessible only in
stars. These can provide constraints on other important parameters for galaxy
evolution, as they trace nucleosynthesis sites complementary to SNe\,II,
which are the main producers of the $\alpha$-elements. Metal-poor dwarf
irregular galaxies have attracted particular attention recently, as they can be
understood as nearby analogues of the basic building-blocks for hierarchical
galaxy formation in the early universe. Spectra of A-type supergiants in the dIrr galaxies 
SMC, NGC\,6822, WLM, Sextans A and GR8 have been obtained with {\sc Vlt/Uves} 
and Keck/{\sc Hires} (Venn~1999; Venn \etal\ 2001, 2003; Kaufer \etal\
2003), requiring exposures of several hours each. At metallicities down to
$\sim$0.05$\times$solar these objects are among the most 
metal-poor massive stars analysed so far. Mean $\alpha$-element abundance
(an average of O, Mg and Si) to iron abundance ratios for the sample stars
as a function of metallicity, [Fe/H],
are compared to Galactic disk and halo stars of similar metallicity in
Fig.~\ref{dirrabundances}. The [$\alpha$/Fe] ratios turn out to be roughly
solar in these systems, indicating a similar contribution of SNe Ia and II
to the chemical evolution as for the young Galactic star
population, thus lacking the $\alpha$-enhancement characteristic of old
star populations. Global lower star formation rates than in the solar neighbourhood 
lead to the lower present-day [Fe/H] of the dIrr galaxies.

\section{{\ldots} and Beyond}
The Local Group provides us with a dozen star-forming galaxies, among those three
giant spirals, where detailed high-resolution studies of individual luminous
stars, primarily BA-type supergiants, are feasible. However, this impressive
laboratory is still insufficient for a comprehensive study of galaxy
formation and evolution. The step beyond the Local Group has to be taken in
order to investigate all the actively star-forming systems along the Hubble sequence 
in clusters and other groups of galaxies, and in the field population. 

\begin{figure}[ht!]
\rule{.7cm}{0cm}\includegraphics[width=.85\linewidth]{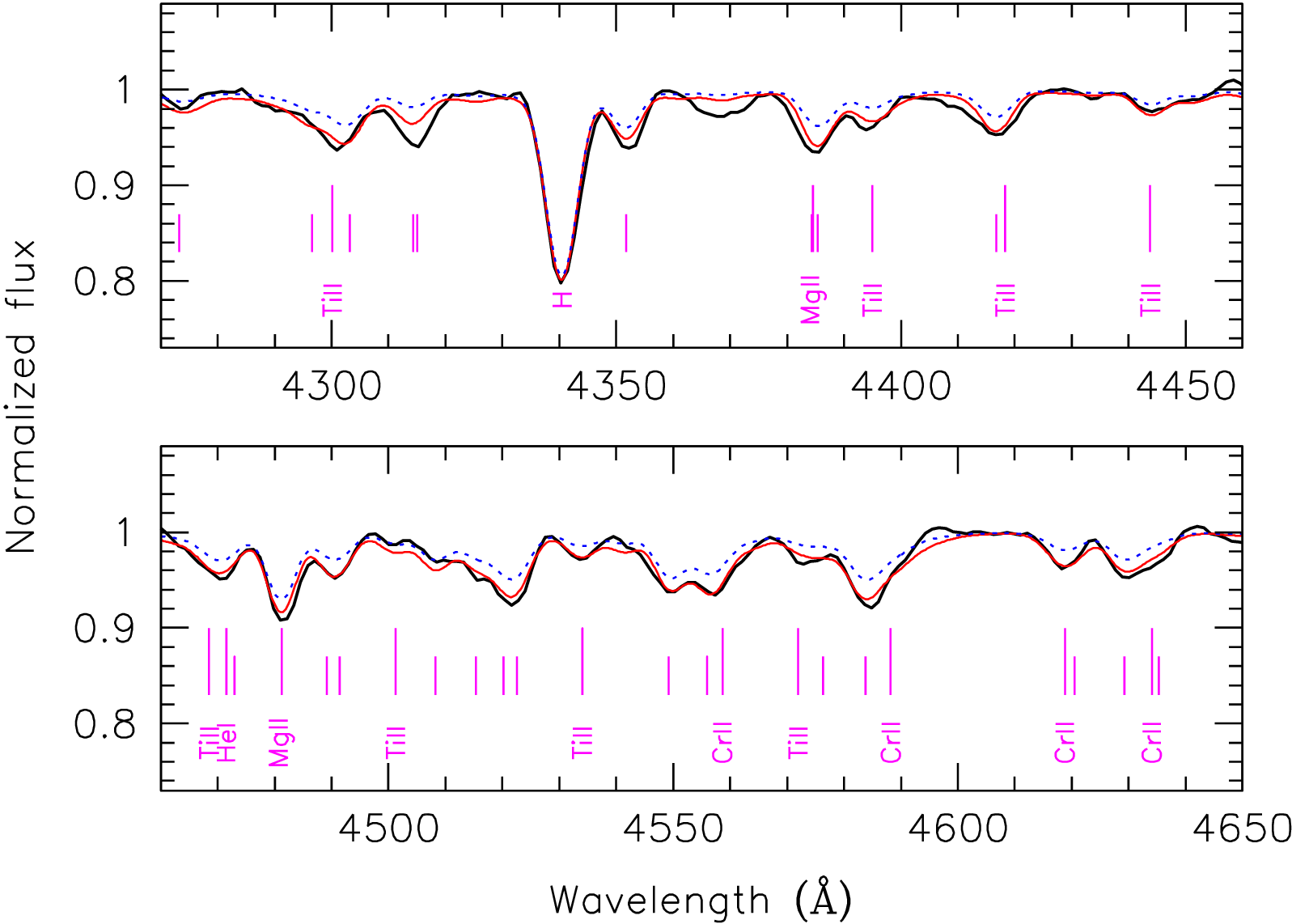}
\caption{Comparison of the medium-resolution spectrum (thick full line) of a 
NGC\,3621 A-type 
supergiant ($V$\,$=$\,20.47\,mag) with model predictions ($T_{\rm eff}$\,$=$\,9000\,K,
$\log g$\,$=$\,1.05) at 0.5 (thin line) and 0.2\,$\times$\,solar metallicity
(dotted line). Line identification is provided (Fe\,{\sc ii} lines are
indicated by the shorter marks). A total of 19 objects in the range
$V$\,$\approx$\,20--22\,mag were observed using the 
multi-object capability of FORS1 on the VLT, of which 10 could be identified
as supergiants, requiring a total integration time of 10.7\,hr.}
\label{extragalstar}
\end{figure}

We have initiated a project to investigate the blue supergiant
populations in nearby galaxies beyond the Local Group, using the {\sc Fors}
multi-object spectrograph on the {\sc Vlt}. Quantitative spectroscopy of
individual stars out to distances of $\sim$7\,Mpc has been performed for the first
time, estimating stellar parameters, chemical abundances, reddening,
extinction and stellar wind properties. The first
target was the field galaxy NGC\,3621 (Bresolin \etal\ 2001). 
Medium-resolution spectra require spectrum synthesis techniques
for the modelling of the entire spectral range in order to facilitate
a closer analysis. While detailed abundance analyses are hampered
by line blending, estimates (within roughly 0.2\,dex) are still possible,
taking full advantage of the methods presented in Sect.~2, as
demonstrated in Fig.~\ref{extragalstar}. The spectrum of the object investigated is 
well fitted with a chemical composition comparable to that of the LMC, i.e.
$\sim$0.5$\times$solar, while a significantly lower metallicity, like that
of the SMC can confidently be ruled out. 

In the second target, the Sculptor Group galaxy NGC\,300, four 
{\sc Fors} fields have been observed, yielding spectra of 62 blue
supergiants. First steps in order to constrain the
NGC\,300 elemental abundance gradients and the internal reddening 
have been undertaken (Bresolin \etal\ 2002; Urbaneja \etal\ 2003).
This will allow, in addition to the topics already mentioned, to address the
influence of two major sources of systematic uncertainty on the Cepheid
period-luminosity relation.



\begin{discussion}
\discuss{Dworetsky}{What values did
you derive for the microturbulent velocity parameter?}

\discuss{Przybilla}{The microturbulent velocities apparently correlate with
luminosity class. At LC Ib typical values are $\sim$4\,km\,s$^{-1}$, increasing
up to $\sim$10\,km\,s$^{-1}$ for objects close to the Eddington limit.
The microturbulent velocities remain sub-sonic in all cases.}

\end{discussion}

\end{document}